\documentclass[a4paper]{jpconf}
\usepackage{graphicx}
\usepackage{hyperref}
\usepackage{amsmath}
\usepackage{amsfonts}
\usepackage{amssymb}
\usepackage{appendix}
\usepackage{pdfpages}
\usepackage{wrapfig}
\usepackage{multirow}
\usepackage{mathrsfs}
\usepackage{epstopdf}
\usepackage{slashed}
\usepackage{soul}
\usepackage{mathtools}
\usepackage{float}
\usepackage{wrapfig}
\usepackage[left=2.5cm,right=2.5cm,top=4cm,bottom=2.7cm]{geometry}
\def\be{\begin{equation}}
\def\ee{\end{equation}}
\def\bea{\begin{eqnarray}}
\def\eea{\end{eqnarray}}
\def\gsim{\ \rlap{\raise 2pt\hbox{$>$}}{\lower 2pt \hbox{$\sim$}}\ }
\def\lsim{\ \rlap{\raise 2pt\hbox{$<$}}{\lower 2pt \hbox{$\sim$}}\ }
\def\dslash{\kern-4pt \not{\hbox{\kern-2pt $\partial$}}}
\def\pslash{\not{\hbox{\kern-2pt p}}}

\newcommand{\dcp}{\delta_{CP}}
\newcommand{\nova}{NO$\nu$A\ }

\begin{document}
\title{The Physics of antineutrinos in DUNE and resolution of octant degeneracy}
%%%%%%%%%%%%
\author{Newton Nath$^{1,2}$, Monojit Ghosh$^{1}$ and Srubabati Goswami$^{1}$}
\address{$^{1}$Physical Research Laboratory, Navrangpura, Ahmedabad 380 009, India, \\
$^{2}$Indian Institute of Technology, Gandhinagar, Ahmedabad--382 424, India}
\ead{newton@prl.res.in}
%%%%%%%%%%%%
%\author{M Ghosh}
%\address{Physical Research Laboratory, Navrangpura,Ahmedabad 380 009, India}
%%%%%%%%%%%%%  
%\author{S Goswami}
%\address{Physical Research Laboratory, Navrangpura,Ahmedabad 380 009, India}

\begin{abstract}
We study the capability of the DUNE experiment, which will be the first beam based experiment  with a wide band flux profile, to  uncover the octant of the leptonic mixing angle $\theta_{23}$ (i.e., $\theta_{23}$ is $< 45^\circ$ or $>45^\circ$). In this work, we find that for the DUNE baseline of 1300 km, due  to enhanced matter effect, the neutrino and antineutrino probabilities are different  which creates a tension in the case of combined runs  because of which octant sensitivity  also can come from disappearance channel. In view of this, we  study the physics of antineutrinos in DUNE and explore the role of antineutrinos run that is required to resolve the octant degeneracy at a certain confidence levels.
\end{abstract}

\section{Introduction}

The fact that neutrinos have tiny non-zero mass is well established from the various neutrino oscillation experiments and the study of neutrino physics has entered  the precision era.  The standard 3-flavor neutrino oscillation scenario has six key parameters and these are,  two $(mass)^{2}$ differences ($\Delta m^2_{i1}, i=2,3$), three mixing angles ($\theta_{ij}, j>i=1,2,3$)  and the CP phase $\delta_{CP}$. Global analysis of neutrino oscillation data provide the best-fit values and $3\sigma$ ranges of these parameters and are given in  \cite{Gonzalez-Garcia:2014bfa,global_fogli,global_valle}.
The remaining unknown in neutrino oscillation physics  are (a) the sign of $|\Delta m^2_{31}|$($\Delta m^2_{31} > 0$, known as normal hierarchy (NH) or $\Delta m^2_{31} < 0$, known as inverted hierarchy (IH)),  (b) the octant of $\theta_{23}$( $\theta_{23} < 45^\circ$, known as lower octant (LO) or $\theta_{23} > 45^\circ$,  known as higher octant (HO)) and  (c) the CP phase $\dcp$. Information on these unknown parameters can come from the currently running long baseline experiments like T2K \cite{t2k_dcphierocthint} and \nova \cite{Adamson:2016tbq,Adamson:2016xxw}. But the main problems which these experiments have to overcome is the issue of parameter degeneracies by which it is meant that different set of parameters can provide equally good fit to the data. 

In this letter our main focus is on the resolution of  octant degeneracy using next generation superbeam experiment DUNE. Currently, neutrino oscillation experiment, like, T2K gives the most accurate  measurements of the parameter $\theta_{23}$. The prime channel for this parameter is the disappearance channel $P_{\mu \mu}$. For baselines shorter than 1000 km this channel has an intrinsic octant degeneracy because of the presence of $\sin^2 2 \theta_{23} $ in the leading order term of the probability expression.  Whereas the probability expression of $P_{\mu e}$ channel depends on the combination of $\sin^2 \theta_{23} \sin^2 2\theta_{13}$ and hence does not suffer from intrinsic degeneracy but has generalized degeneracies which were addressed in \cite{Barger:2001yr, Minakata:2001qm, BurguetCastell:2002qx, Coloma:2014kca, Ghosh:2015ena}. Here,  our main goal is to understand the role of antineutrinos in enhancing  the octant sensitivity of DUNE experiment. It was observed that the main complication that arises in octant determination is due to the unknown value of $\dcp$ in the subleading terms of $P_{\mu e}$ channel which gives rise to octant-$\dcp$ degeneracy.  It was discussed  in \cite{Huber:2009cw, Minakata:2003wq} that combining the reactor measurement of $\theta_{13}$ with the accelerator data will be beneficial for extraction of information on the octant value from the $P_{\mu e}$ channel. 
%Thus, the accurate measurement of 1-3 mixing angle from the reactor experiments is expected to increase the octant sensitivity. 
The combined measurement of $P_{\mu \mu}$ channel  with $P_{\mu e}$ channel in 
long baseline experiments can also be helpful in determining the octant degeneracy 
because of the non-identical functional dependence of the two probabilities 
on $\theta_{23}$. We present the results of the octant sensitivity using various combinations of ($ \nu $ and $ \overline{\nu} $) run  as a function of true $\dcp$ for fixed values of 
true $\theta_{23}$. In addition, we also present the precision of $\theta_{23}$ in (true:$\theta_{23}$ - test:$\theta_{23}$) plane.  The details of numerical simulation and experimental specifications that we have considered are given in \cite{Nath:2015kjg} and the references there in.

\section{Result}
%%%%%%%%%%%%%%%%%%%%%%%%%%%%%%%%%%%%%%%%%%%%

In this section we discuss the octant sensitivity of DUNE experiment. We get four possibilities of (hierarchy$ - $octant) depending on the true parameters and which are, NH-LO, NH-HO, IH-LO and IH-HO. Here, we only discuss the cases NH-LO and IH-LO. Details and the remaining 2 cases are describe in \cite{Nath:2015kjg}. Here, in the first column dark-blue curves represents True:NH - Test:NH and magenta curves represents True:NH - Test:IH  and in the second column dark-blue curves represents True:IH - Test:IH. In both the plots the horizontal yellow line represents the octant sensitivity at 3$ \sigma $ C.L.  Representative true value of $ \theta_{23} $ for this two plots are 39$^\circ$(LO).

\begin{figure}[H]
%\vspace{-1.7cm} 
        \begin{tabular}{lr}
                \hspace*{0.35in} 
                \includegraphics[width=3in]{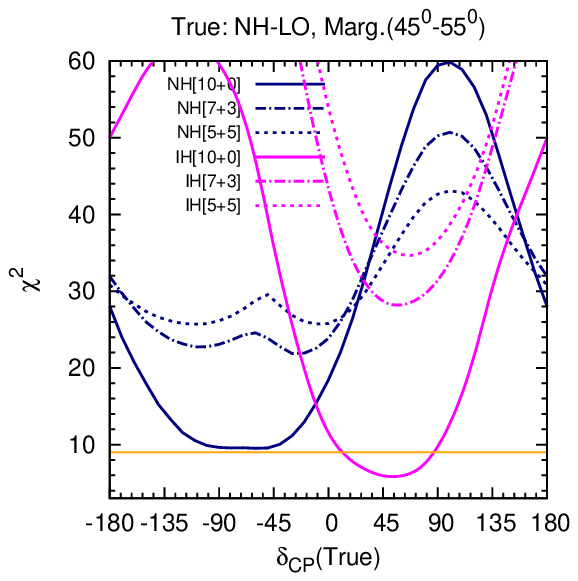}
                & 
                \hspace*{-0.9in}
               \includegraphics[width=3in]{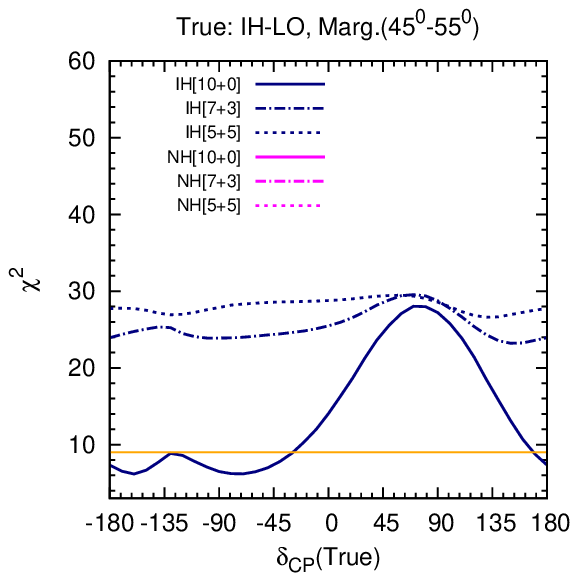}
%               \\
%                \hspace*{-0.35in} 
%                \includegraphics[width=0.5\textwidth]{lbne_oct_chisq_tr_ih_nh_lo.eps}
%                & 
%                \hspace*{-1.3in}
%                \includegraphics[width=0.5\textwidth]{lbne_oct_chisq_tr_ih_nh_ho.eps}
        \end{tabular}
\vspace{-0.5cm}        
\caption{This figures represent octant discovery $ \chi^{2} $ for DUNE. Here, left (right) column is for true NH (IH) and for both the figures  true($ \theta_{23} = 39^\circ$ ) and test($ \theta_{23}$) is marginalized over ($ 45^\circ $ to $ 55^\circ $) and the labels
NH and IH inside the plots signifies test hierarchy.}
\label{fig:oct_discovery}
\end{figure}

 Let consider the case for NH-LO, from the left column of fig.(\ref{fig:oct_discovery}) we can see that [10+0] year $ \nu $ run of DUNE can resolve octant degeneracy at 3$ \sigma $ C.L.(solid blue curve) if the hierarchy is known. The main point to notice here is that,  though only neutrino run suffers from octant degeneracy, still we get octant sensitivity around 3$ \sigma $. This is one of the special features of the on axis beam based neutrino oscillation experiment where the degeneracy does not exist over the complete energy range and  one can still have  some octant sensitivity only from the neutrino channel. However, storyline changes  if the hierarchy is unknown then there is a   wrong hierarchy(WH)-wrong octant(WO) solutions in the region $ 9^\circ < \dcp < 90^\circ $.  We see that once antineutrino run added, say [7+3] years of ($ \nu + \overline{\nu}  $) run, octant degeneracy can be resolved with more than 4$ \sigma $ C.L. even without the 
knowledge of the true hierarchy for all values of $\dcp$.
%%%%%%%%%%%%%%%%%%%%%%%%%%%%%%%%%%%%%%%%%%%%
%$$$$$$$$$$$$$$$$$$$$$$$$$$$$$$$
In the case of IH-LO as shown in the  right column of fig.(\ref{fig:oct_discovery})  when  $ -180^\circ < \dcp < 0^\circ $ only neutrino suffers from degeneracy whereas addition of antineutrino run enhances the sensitivity because they  do not suffer from octant degeneracy \footnote{Note that RHS of fig.(\ref{fig:oct_discovery}) does not contain any magenta curve because of the absence of WH-WO solution for IH-LO.} 
and hence % It is seen that for IH, because of the enhancement of the antineutrino probability due to  matter effect,
  a large octant sensitive contribution to the $\chi^2$ is added. These combinations of hierarchy$ - $octant can resolve octant degeneracy with  5$ \sigma $ sensitivity with (5+5) years of ($ \nu + \overline{\nu} $) run for any value of $ \delta_{CP} $.
%%%%%%%%%%%%%%%%%%%%%%%%%%%%%%%%%%%%%%%%%%%%
\begin{figure}[H]
        \begin{tabular}{lr}
                \hspace*{0.35in}
                \includegraphics[width=3in]{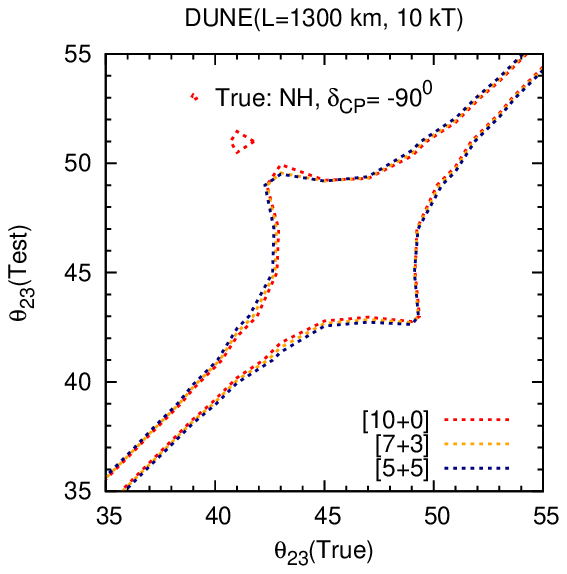}
                \hspace*{-1.0in}
                 \includegraphics[width=3in]{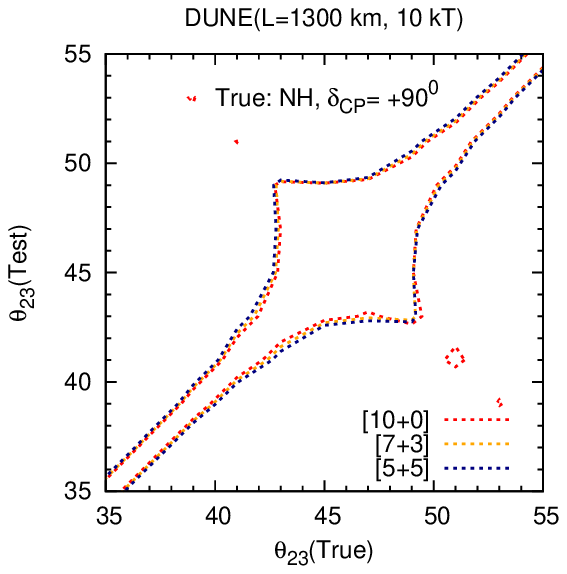}          
        \end{tabular}
\vspace{-0.5cm}
\caption{Here, the plots are $ \theta_{23} $ precision plots of DUNE in True($ \theta_{23} $) -  Test($ \theta_{23} $) plane at 3$ \sigma $  C. L. Here both the columns are for NH and left(right) column is for true:$ \delta_{CP} $ = $ - 90^\circ $($ +90^\circ $).}
\label{fig:th23_precision} 
\end{figure}
%##############################################################
%$$$$$$$$$$$$$$$$$$$$$$$$$$$$$$$$$$$$$$$$$$$$$$$$$$$$$$$$$$$$$$$$
In Fig. \ref{fig:th23_precision} we plot the $3 \sigma$  precision contours in the (true:$\theta_{23}$ - test:$\theta_{23}$) plane for true $\dcp = - 90^\circ$ (left column) and $\dcp = + 90^\circ$ (right column) for normal hierarchy (for inverted hierarchy see the discussion on \cite{Nath:2015kjg}). Both these figures demonstrate 
the relation between precision of $\theta_{23}$ and octant degeneracy. From these plots we see that only for neutrino run (i.e (10+0)) there are other allowed values of $\theta_{23}$ apart from the true value, if $\theta_{23} \in$ LO (HO) at $\dcp=-90^\circ (+90^\circ)$ . 
This happens because of the existence of octant degeneracy. 
As we have  seen from fig.(\ref{fig:oct_discovery}) that for $\dcp=-90^\circ (+90^\circ)$, neutrinos suffer from octant degeneracy in LO (HO)   and this in turn
affects the precision of $\theta_{23}$ which is clearly seen
from this figures. Here, we see that the addition of antineutrinos help to improve the precision in both [7+3] and [5+5] curves and both the curves give almost similar precision of $\theta_{23}$. But as one approaches the maximal value of $\theta_{23}$ from non-maximal value, the precision becomes worse due to the difficulty in determining the octant around  $\theta_{23}$ maximal.

In conclusion, we have explored the significant role of antineutrinos in 
giving an enhanced  octant sensitivity in next generation superbeam experiment DUNE which has a 1300 km baseline  with a broad-band beam.
% We focus on the importance of antineutrino run in resolving octant degeneracy. 
In our study, we observe that although for some specific parameters only neutrino run can  provide $3\sigma$ octant sensitivity,  overall combined ($ \nu + \overline{\nu}  $) run gives better sensitivity. 

%%%%%%%%%%%%%%%%%%%%%%%%%%%%%%%%%%%%%%%%%%%%%%
%\bibliography{neutosc_new}{}
\end{document}